# Single molecule Raman spectroscopy and local work function fluctuations

Gilad Haran

Chemical Physics Department

Weizmann Institute of Science

Rehovot 76100, Israel

e-mail: gilad.haran@weizmann.ac.il

**Abstract:**

Single molecule Raman spectroscopy provides information on individual molecules with vibrational-level resolution. The unique mechanisms leading to the huge Raman cross-section enhancement necessary for single molecule sensitivity are under intense investigation in several laboratories. We recently analyzed large spectral fluctuations in single molecule spectra of rhodamine 6G on silver surfaces (A. Weiss and G. Haran, J. Phys. Chem. B (2001), 105, 12348-12354). The appearance of the fluctuations in two particular vibrational bands, and their dependence on several parameters, suggested that they originate in a charge transfer interaction of an adsorbed molecule with the surface. We argued that the fluctuations are due to variations of the local work function at the position of the molecule. In the current paper the fluctuations are further analyzed in terms of the intensity ratio between a fluctuating and a quiescent band, and it is found that the distribution of this ratio is independent of laser power, unlike the correlation time of the fluctuations. We show that a simple model, based on the energetics of charge transfer, can be used to extract the local work function distribution from the intensity ratio distribution. In a second experiment, single molecule spectra are collected from colloids immersed in water and in glycerol and a threefold decrease in fluctuation rate is found in the more viscous fluid. This indicates that surface dynamics are indeed responsible for the fluctuations, involving the motion of the adsorbed molecule and possibly also that of surface silver atoms around it.

The spectroscopy of single molecules opens up new opportunities for the analysis of molecular dynamics. Much of the single molecule spectroscopy field is involved with fluorescence measurements. However, fluorescence is limited as a source of spectroscopic information on molecules, especially at room temperatures, where fluorescence spectral lines are broad and featureless. Thus, Raman spectroscopy is a very welcome addition to the arsenal of methods aimed at studying individual molecules. The large scattering cross section required for the detection of Raman signals from single molecules is attained through surface enhancement. A judicious application of this technique must rely on a thorough understanding of the mechanism of this enhancement. Surprisingly, even after more than 25 years of intense research into the phenomenon of surface-enhanced Raman scattering (SERS), many basic questions still remain unexplained. In fact, single molecule studies might reflect back on the SERS field by teaching us new facts and providing fresh explanations.

This paper is opened with a discussion of current thinking on the microscopic processes leading to the huge scattering enhancement observed in single molecule Raman measurements. Our own recent work, in which we studied fluctuations in Raman spectra of individual rhodamine 6G (R6G) molecules on silver colloidal nanoparticles, is reviewed, and the various possibilities for explaining the experimental results are delineated. A particular emphasis is given to the role of charge transfer interactions in the Raman process. It is shown that the distribution of spectral fluctuations can be used to obtain information about the range of local work function (LWF) values sampled by a molecule. Finally, more evidence for the role of molecular motion in the spectral fluctuations is drawn from experiments in which the viscosity of the medium in which the colloids are immersed is varied.

**Enhancement mechanisms in single-molecule SERS (smSERS)**

SmSERS, first observed in 1997 with R6G molecules [1,2], has been since observed with several different molecules [3-7], including proteins [8-10] and DNA [11]. Single molecules of R6G exhibit a particularly large scattering cross section on silver colloids under resonance conditions, up to $10^{-13}$ cm$^2$ [12]. This large cross section has in fact been noted in bulk experiments as far back as 1984 [13], and involves enhancement by some ten orders of magnitude. What are the microscopic mechanisms behind such a phenomenal enhancement factor?

As is well-known, the main and foremost contribution to SERS comes from the enhancement of electromagnetic (EM) fields close to the surface through interaction with surface plasmon excitations (for reviews see 14,15). The local EM field depends on the microscopic shape of the metallic surface, such as the presence of sharp edges. Calculations by several authors [16-21] suggest that the EM field generated inside junctions between nanoparticles forming small aggregates may lead to a giant enhancement of Raman scattering, up to a factor of $10^{11}$. The groups of Käll [8] and Brus [22] used AFM to show that single molecule signals come from small aggregates containing two or several nanocrystals, implying that 'enhanced' molecules indeed reside within interparticle junctions. Recent work from the Brus lab showed that the depolarization of the overall Raman signal depends on the direction of the excitation field [23]. This is exactly the effect expected when the Raman scattering is localized in a junction, which is an anisotropic structure.

A second (and smaller) contribution to the enhancement of Raman scattering is a specific interaction of the adsorbed molecule with the metal surface [15,24], which leads to charge transfer from the molecule into the empty levels on the metal surface or from occupied surface levels to the molecule. This electron transfer can in fact be viewed as an electronic excitation of the coupled molecule-surface system, and should therefore lead to the appearance of a new band in the electronic spectrum of the molecule. Charge transfer (CT) bands have been demonstrated in optical [25] and electron energy loss spectra [26] of adsorbed molecules, and their relevance to SERS was discussed by Persson [27] and others. The contribution of CT transitions to SERS enhancement can be seen as a resonance Raman process [28], and since the energy of a CT band depends on surface potential, the Raman enhancement can be modified by tuning this potential. This has been verified in many systems, the classical examples being those of pyridine and pyrazine [29,30]. The CT enhancement mechanism is restricted, by its nature, to molecules adsorbed directly on the metal, as opposed to the EM effect which extends a certain distance beyond the surface. The role of the CT mechanism in single molecule Raman scattering of R6G was discussed by several groups [3,11,12], and is the subject of this paper too.

**Charge transfer resonance in single-molecule SERS of rhodamine 6G**

Strong temporal fluctuations in specific bands were seen in the Raman spectra of individual R6G molecules adsorbed on surface-immobilized silver colloidal silver particles (Figure 1A), and were the basis for our assertion that the CT mechanism is operative in the case of these molecules [12]. Recently we also measured similar fluctuations from R6G molecules adsorbed on silver island films (Figure 1B). In both cases the spectra were taken with a home-built Raman spectrometer equipped with a 532 nm laser, a 0.15 m spectrograph and a back-illuminated CCD camera [12]. The fluctuations, on a time scale of many seconds, appear in two bending bands, at 614 and 774 cm$^{-1}$, and are uncorrelated with the much weaker fluctuations of other bands in the spectrum (Figure 2). The lack of correlation between various parts of the spectrum suggests that it is unlikely to attribute the fluctuations to orientational motion of the molecules. We therefore proposed that the two fluctuating bands are resonance-enhanced by coupling to a transition that the rest of the bands are not coupled to. This transition can be identified as a molecule-surface CT transition.

Evidence for the occurrence of this CT transition is found in the work of Hildebrandt and Stockburger [13]. These authors measured the SERS excitation profiles of various bands of R6G. The excitation spectrum they presented for the 774 cm$^{-1}$ band showed an extra shoulder not seen in excitation spectra of other bands. This shoulder, appearing at energy of 2.2 eV, could well be due to a CT transition.

To gain further understanding of the mechanism of the fluctuations, it is useful to discuss the energetics involved in the CT transition between an adsorbed molecule and a metal surface. The CT energy, in the case where electron transfer occurs from the surface to the molecule, can be written as follows:

1) $E_{CT} = \phi + E_{LUMO} - E_{vac} - E_{int}$

where $\phi$ is the work function of the metal, $E_{LUMO}$ is the absolute energy of the lowest unoccupied molecular orbital (LUMO) of the molecule, $E_{vac}$ is the vacuum energy level and $E_{int}$ is the energy due to interaction of the charge on the molecule with its image in the metal [15]. All energies are measured relative to the Fermi level of the metal. A similar expression can be written when electron transfer occurs from the molecule to the metal. In the case of the positively-charged R6G it is expected that the direction of CT will be from metal to molecule. A diagram of the various energy levels involved in the CT process of a R6G molecule on a silver colloid is shown in Figure 3.

The factor most likely to fluctuate in equation 1 is the work function. Indeed, local variations in the work function have been extensively discussed in the literature [31]. The physics of such variations goes back to Smoluchowski [32], who noted that the electron density at the surface of a metal cannot follow sharp edges, such as steps. The smoothing of the electron density leads to a reduction of the electrical potential, which in turn lowers the work function at the position of the sharp edge. The concept of the local work function (LWF) has been introduced in order to allow a quantitative description of such variations along the surface. There are several experimental methods that can probe the LWF [31,33].

In our previous work we showed that the spectral fluctuations depend on laser power through a non-thermal mechanism. We further showed that the salt concentration in the solution in which the silver colloids are immersed also modulates the rate of fluctuations. We suggested that molecular motion, triggered by light, is involved in the fluctuations by allowing each molecule to sample various LWF values on the surface. More recently we argued that dynamic relaxation of surface roughness can also be modulated by salt concentration and possibly by light [34]. The coupling of salt concentration and electromagnetic field to surface dynamics may come through their effect on the surface tension.

**Spectral fluctuations and the LWF distribution**

While the mechanism by which the LWF changes at the adsorption site of a molecule is not clear yet, and can be attributed to molecular diffusion but also to surface dynamics, we can still build a simple model that will allow us to translate spectral fluctuations to LWF fluctuations. To do this, we focus here on spectral changes in one vibrational band of R6G, the C-C ring in-plane bend at 614 cm$^{-1}$ [35-37], and calculate the ratio of the time-dependent intensity of this band to the intensity of one of the bands that do not show strong fluctuations, the aromatic C–C stretch at 1650 cm$^{-1}$. This ratio, as obtained from one molecule, is plotted in Figure 4A. For comparison, the ratio between the two correlated bands at 1575 cm$^{-1}$ and 1650 cm$^{-1}$ is shown in Figure 4B. The distributions of the fluctuations in the ratio between the intensities of the 614 cm$^{-1}$ and 1650 cm$^{-1}$ bands, obtained at two different laser powers, 6 and 24 Watt/cm$^2$, are plotted in Figure 5A. Each distribution was calculated from data similar to that of Figure 4A, collected from ~20 molecules (the molecules were adsorbed on silver nanoparticles which were then immobilized on a glass surface).

As noted above, it has been established that the rate of fluctuations depends linearly on laser power. It is important to test whether the amplitude of the fluctuations also depends on this variable. In other words- does the increase in laser power lead to sampling of a larger set of LWF values, or is it only the rate of sampling that changes with laser power? Figure 5A shows that the distributions obtained at 6 and 24 Watt/cm$^2$ are very similar, and the small differences between them are probably due to insufficient sampling. This indicates that the range of values of the LWF sampled in the experiment does not depend on laser power. We therefore continue the analysis with a distribution which is the average of the two distributions of Figure 5A.

In order to invert this distribution and extract from it the distribution of LWF values, we first normalize all band intensity ratio values (labeled R) by the largest possible ratio, $R_{max}$, for which a value of 6 is chosen (the fit described below does not depend significantly on the exact value chosen). We assume that the dependence of the band intensity ratio on $E_{CT}$ can be described as a Lorenzian-shaped resonance:

2) $$\overline{R}(\Delta E) = \frac{\beta^2}{\Delta E^2 + \beta^2}$$

In this equation $\overline{R} = R/R_{max}$ and $\Delta E$ is the fluctuation of $E_{CT}$ away from resonance with the laser frequency, 2.35 eV. $\beta$ is a parameter which determines the width of the resonance peak. $\beta$ can be calculated using two parameters known from the ensemble spectrum: the average shift of the CT band from resonance with the laser, $\Delta E_0$, and the average value of the normalized band intensity ratio, $\bar{R}_{av}$. The first parameter is estimated from the SERS excitation profiles of Hildebrandt and Stockburger [13] to be 0.15 eV. The second parameter is directly calculated from the experimental data to be 0.17. From these two parameters we calculate a value of 0.09 eV for $\beta$.

The distribution of the $E_{CT}$ fluctuation is described as a Gaussian function:

3) $P(\Delta E) = (2\pi\sigma^2)^{-1/2} \exp(-\frac{(\Delta E - \Delta E_0)^2}{2\sigma^2})$

where $\sigma$ is the standard deviation of the distribution. We can invert equation 1 to get an expression for $\Delta E(\bar{R})$ and introduce that expression in equation 3 to obtain the normalized distribution as a function of $\bar{R}$, $P(\bar{R})$, in which $\bar{R}$ may run from 0 to 1.

4) $P(\bar{R}) = \frac{1}{2}\beta(2\pi\sigma^2)^{-1/2}(\bar{R}^{-1}-1)^{-1/2}\bar{R}^{-2}\left[\exp(-\frac{(\beta(\bar{R}^{-1}-1)^{1/2}-\Delta E_0)^2}{2\sigma^2}) + \exp(-\frac{(\beta(\bar{R}^{-1}-1)^{1/2}+\Delta E_0)^2}{2\sigma^2})\right]$

Equation 4 contains only one unknown parameter, $\sigma$ (but see below). The green curve in Figure 5B is a least-squares fit to the ratio distribution using this equation. From the fit we find that $\sigma$ takes a value of 0.06 eV. The small value of $\sigma$ that fits the experimental data tells us that the distribution of LWF values sampled by R6G molecules is rather narrow. On one hand, the sensitivity to such small variations in the LWF attests to the forte of single molecule Raman measurements. On the other hand, it also teaches us that the dynamics involved in the Raman process are actually rather modest in scale. Indeed, literature data shows that the LWF on certain metals can vary by as much as 1 eV [31]. It should be noted though that the value of $\sigma$ obtained in the fit depends on the value selected for $\Delta E_0$, and becomes larger as $\Delta E_0$ is increased. As indicated above, we have only an indirect estimation for this value. It would be useful to find a method to verify independently the position of the CT resonance of R6G.

**Effect of liquid viscosity on spectral fluctuations**

If the mechanism of the spectral fluctuations involves translational dynamics of either the R6G molecules or of silver atoms, then in principle it should be possible to influence these fluctuations by changing the viscosity of the liquid in which the colloids are immersed. In order to test this assertion we collected sequences of spectra from single R6G molecules adsorbed on colloids, which were immobilized on a glass surface and immersed in either water or in glycerol. Spectral correlation functions were calculated from these spectra, following the scheme in reference 12. The spectra were all normalized (to get rid of the total intensity fluctuation contribution), and a correlation function was calculated for each molecule according to equation 5,

$$5)\ C(\tau) = \sum_{\bar{v}} \left( \langle I_{\bar{v}}(t) \cdot I_{\bar{v}}(t+\tau) \rangle - \langle I_{\bar{v}}(t) \rangle^2 \right)$$

in which $I_{\bar{v}}(t)$ is the intensity at wavenumber $\bar{v}$ at time t, including the diffuse background, and the angular brackets denote temporal averaging. Inclusion of the background is not only a numerical convenience, but is justified by the strong relation of background and spectral fluctuations, as discussed before. Correlation functions were first calculated for each molecule and then averaged over all molecules belonging to the same series. The calculated correlation functions are shown in Figure 6. The correlation time (estimated e.g. as $t_{1/2}$) is approximately three times faster in water than in glycerol. This points out to the involvement of translational diffusion in the mechanism of the fluctuations. The source of this diffusion is likely to be motion of R6G molecules on the surface, but can also involve motion of silver atoms involved in surface relaxation dynamics, as discussed above. Both types of motion might be sensitive to the viscosity of the solution covering the colloids.

It is pertinent to ask, though, why the effect of glycerol on the fluctuations is not much larger, considering the fact that its viscosity is more than a hundred times larger than that of water, even when a small amount of water is dissolved in it. One possibility is that the interfacial viscosity of glycerol at the silver surface is much lower than bulk viscosity, or that a strong hydration layer remains on the colloids when they are transferred from water to glycerol, so that the interfacial viscosity is more similar to that of water. Another possibility is that factors having to do with surface structure and energetics are more important in determining the physics of the translational diffusion process governing fluctuations than the viscosity of the medium.

**Conclusion**

In this article we focused on the fluctuations in the Raman spectra of individual R6G molecules. The evidence linking these fluctuations to the charge transfer enhancement mechanism of SERS was reviewed. New analysis was presented that allowed us to extract the distribution of local work function from experimental SERS spectra. In addition, the effect of solution viscosity on the fluctuations was established, providing another indication for the importance of surface dynamics in the mechanism of the fluctuations.

The picture arising from all recent studies of smSERS is rather clear. The large enhancement allowing visualization of single molecules probably comes from the huge EM field in junctions between nanoparticles. Surface-molecule charge transfer resonance adds to this enhancement and provides for a specific effect on some of the bands in the SERS spectrum. This effect makes smSERS sensitive to surface dynamics, the nature of which is still not fully understood. Future studies in our lab will address the open questions, such as the 'chemical' nature of the adsorption sites on the surface, by comparing single molecule Raman scattering with several different molecules as well as different substrates.

**Acknowledgement.** I would like to thank Tali Amitay, Timur Shegai and Amir Weiss for performing some of the measurements reported in this paper.

**Figure Legends**

**Figure 1:** Substrates used in single molecule Raman scattering studies of R6G in our lab. A. Colloidal nanoparticles. Transmission electron micrograph showing ~50 nm silver particles prepared by the Lee-Meisel method [38]. The formation of small aggregates is typical and might be crucial for the huge enhancement factors provided by such particles. B. Island films. AFM image of a silver island film (nominal thickness 5 nm), which was prepared by a standard vacuum deposition technique followed by high-temperature annealing.

**Figure 2:** Two Raman spectra from a time-series obtained from a single molecule of R6G, demonstrating the large variability in the amplitude of the bands at 614 cm$^{-1}$ and 774 cm$^{-1}$.

**Figure 3:** Energy level diagram of R6G on a silver surface. $E_F$- Fermi energy of the metal. $\phi$ – the average bulk work function of silver, 4.6 eV. HOMO and LUMO - highest occupied and lowest unoccupied molecular orbitals of R6G. Placement of the molecular energy levels with respect to the metal levels is based on the assumption that a CT resonance exists at 2.2 eV. Charge transfer from metal to molecule is the likely process for the positively-charged R6G molecule.

**Figure 4:** Temporal variation of ratios between vibrational band intensities in the SERS spectrum of a single R6G molecule. A. $R = I_{614 cm^{-1}} / I_{1650 cm^{-1}}$, B. $R = I_{1575 cm^{-1}} / I_{1650 cm^{-1}}$. The strong modulation of the ratio in A reflects mainly large intensity changes of the bending band at 614 cm$^{-1}$. Indeed, the ratio between the two aromatic C-C stretch bands in B does not show this strong modulation.

**Figure 5**: A. Distributions of the ratio between the 614 cm$^{-1}$ and 1650 cm$^{-1}$ bands, obtained from a series of R6G single molecule spectra at two laser powers, 6 W/cm$^2$ (full symbols) and 24 W/cm$^2$ (empty symbols). The ratio scale is normalized by its maximum value $R_{max}$=6. B. Fit of the model described in the text to the average of the two distributions in A. The average distribution is shown as gray bars, while the fit is shown as a black line.

**Figure 6:** Correlation functions calculated from single molecule spectra. In full line- the correlation function taken with the colloids immersed in water. In dashed line- the correlation function taken with colloids immersed in glycerol. Each curve is the average of ~20 single molecule correlation functions. Straight lines guide the eye to the threefold difference in decay time between the two curves.

Figures

Figure 1

A 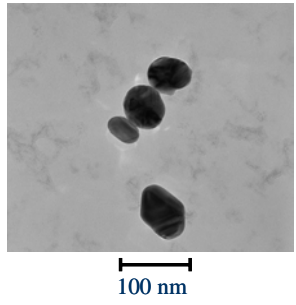 B 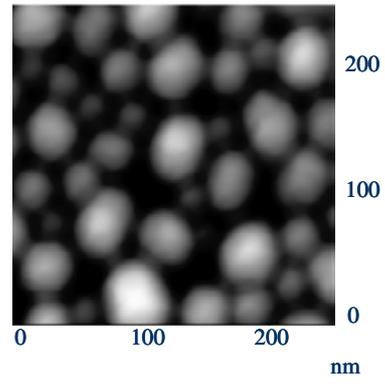

Figure 2

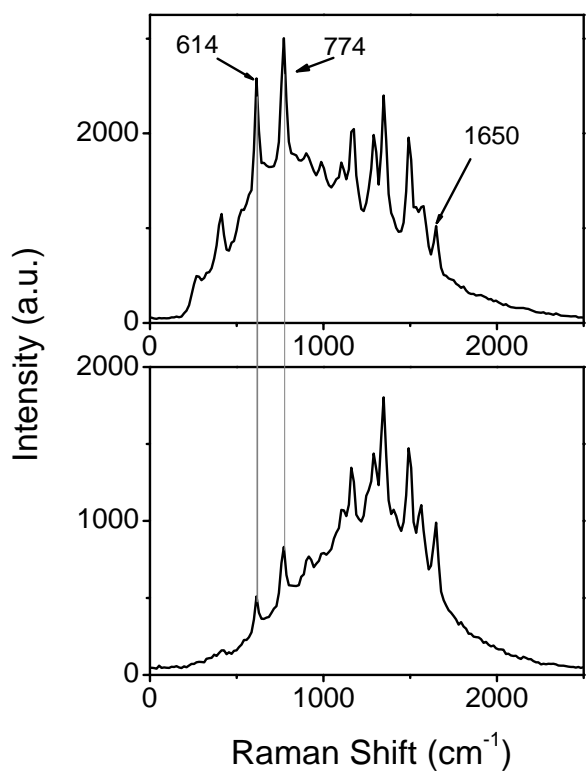

Figure 3

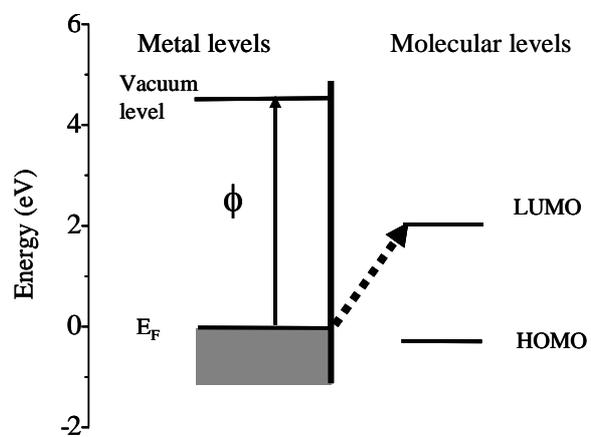

Figure 4

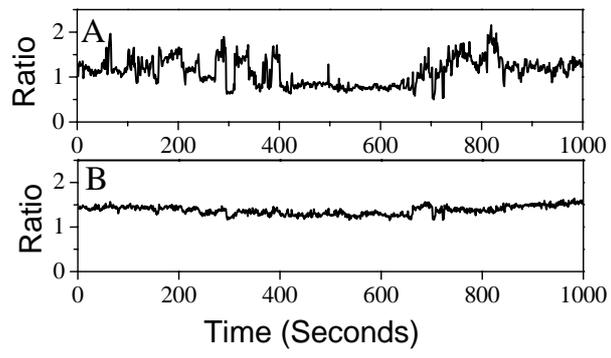

Figure 5

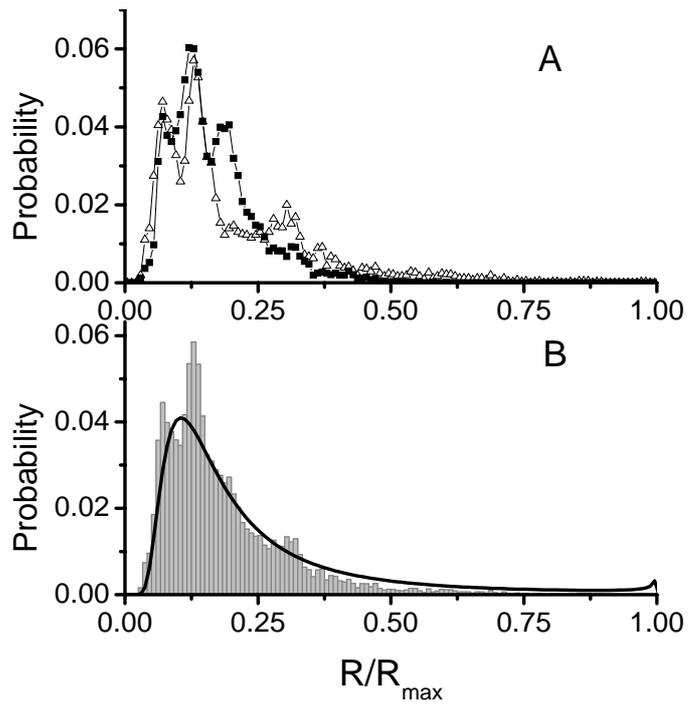

Figure 6

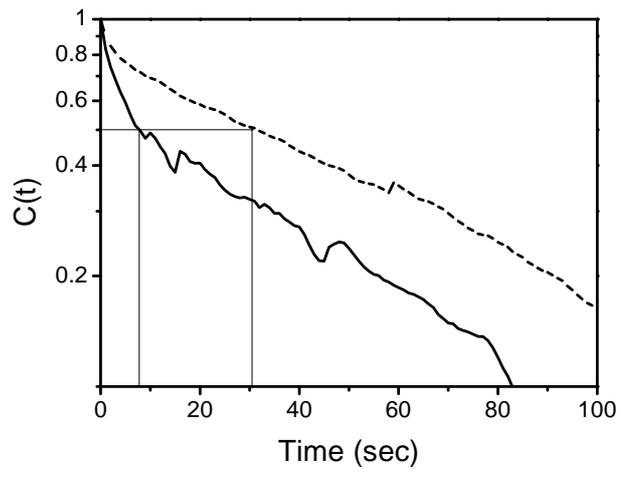